\documentclass[aps,preprint,noshowpacs,superscriptaddress,groupedaddress]{revtex4}

\usepackage{graphicx}
\usepackage{dcolumn}

\begin{document}

\title{Layering transitions in superfluid helium adsorbed on a carbon nanotube mechanical resonator}

\author{Adrien Noury}
\altaffiliation{Present address: Laboratoire Charles Coulomb (L2C), Univ Montpellier, CNRS, Montpellier, France}
\affiliation{ICFO - Institut De Ciencies Fotoniques, The Barcelona Institute of Science and Technology
Mediterranean Technology Park, 08860 Castelldefels (Barcelona), Spain}
\author{Jorge Vergara-Cruz}
\affiliation{ICFO - Institut De Ciencies Fotoniques, The Barcelona Institute of Science and Technology
Mediterranean Technology Park, 08860 Castelldefels (Barcelona), Spain}
\author{Pascal Morfin}
\affiliation{Laboratoire de Physique de l'\'{E}cole normale sup\'{e}rieure, ENS, Universit\'{e} PSL, CNRS, Sorbonne Universit\'{e}, Universit\'{e} Paris-Diderot, Sorbonne Paris Cit\'{e}, Paris, France}
\author{Bernard Pla\c{c}ais}
\affiliation{Laboratoire de Physique de l'\'{E}cole normale sup\'{e}rieure, ENS, Universit\'{e} PSL, CNRS, Sorbonne Universit\'{e}, Universit\'{e} Paris-Diderot, Sorbonne Paris Cit\'{e}, Paris, France}
\author{Maria C. Gordillo}
\affiliation{Departamento de Sistemas F\'{i}sicos, Qu\'{i}micos y Naturales, Universidad Pablo de Olavide
Carretera de Utrera, km 1, E-41013 Sevilla, Spain}
\author{Jordi Boronat}
\affiliation{Departament de F\'{i}sica, Universitat Polit\`{e}cnica de Catalunya
B4-B5 Campus Nord, 08034 Barcelona, Spain}
\author{S\'{e}bastien Balibar}
\affiliation{Laboratoire de Physique de l'\'{E}cole normale sup\'{e}rieure, ENS, Universit\'{e} PSL, CNRS, Sorbonne Universit\'{e}, Universit\'{e} Paris-Diderot, Sorbonne Paris Cit\'{e}, Paris, France}
\author{Adrian Bachtold}
\affiliation{ICFO - Institut De Ciencies Fotoniques, The Barcelona Institute of Science and Technology
Mediterranean Technology Park, 08860 Castelldefels (Barcelona), Spain}


\begin{abstract}

Helium is recognized as a model system for the study of phase transitions. Of particular interest is the superfluid phase in two dimensions. We report measurements on superfluid helium films adsorbed on the surface of a suspended carbon nanotube. We measure the mechanical vibrations of the nanotube to probe the adsorbed helium film. We demonstrate the formation of helium layers up to five atoms thickness. Upon increasing the vapour pressure, we observe layer-by-layer growth with discontinuities in both the number of adsorbed atoms and the speed of sound in the adsorbed film. These hitherto unobserved discontinuities point to a series of first-order layering transitions. Our results show that helium multilayers adsorbed on a nanotube are of unprecedented quality compared to previous works. They pave the way to new  studies of quantized superfluid vortex dynamics on cylindrical surfaces, of the Berezinskii-Kosterlitz-Thouless phase transition in this new geometry, perhaps also to supersolidity in crystalline single layers as predicted in quantum Monte Carlo calculations.

\end{abstract}


\maketitle

When exposing graphite  to a helium vapor at low temperature, a helium  film forms on the graphite surface. Several experiments have shown that the thickness of this film grows layer by layer as a function of the vapor pressure. These layers are one atom thick~\cite{Greywall1991, Greywall1993, Chan1992, Reppy1996, Saunders2017}, and there are successive ``layering transitions'' between layers n and n+1. Clements et al.~\cite{Clements1996,Clements1993} had  predicted that the layering transitions are first order transitions, meaning that the coverage should show a series of sharp discontinuities as a function of the helium pressure.  However, such  discontinuities could not be observed, probably because  previous techniques required large surface areas, which were obtained e.g. by  chemically exfoliating graphite and by subsequently pressing the material into sheets. As a result, the substrate surface was made of small crystalline  platelets of $\sim (10 $nm$)^2$ area~\cite{Reppy1996} and probably contained a sizeable amount of defects including wedges where liquid helium could accumulate. Despite the modest quality of these surfaces, evidences for layer-by-layer growth were reported but without any visible discontinuities \cite{Godfrin1995, Chan1992, Reppy1996, Saunders2017, Menachekanian2019}.

By studying the mechanical resonance of a single wall carbon nanotube (NT) we have  observed discontinuities in  the adsorbed mass as a function of the injection of helium atoms and demonstrated for the first time that the layering transitons are indeed first order. A carbon nanotube is an excellent substrate, since the crystalline quality is high, and such nanotubes can be made free of adsorbed contamination~\cite{Tavernarakis2014}. Our results prove the very high quality of carbon nanotubes as potential substrates for new  original studies like quantized superfluid vortex dynamics on cylindrical surfaces, or the Berezinskii-Kosterlitz-Thouless phase transition in a new geometry, perhaps also of supersolidity in crystalline single helium layers of variable density as predicted by Gordillo et al.\cite{Gordillo2011}.

\begin{figure}[h!]
\centering
\includegraphics[scale=0.4, trim = 0cm 0cm 0cm 0cm, clip=true]{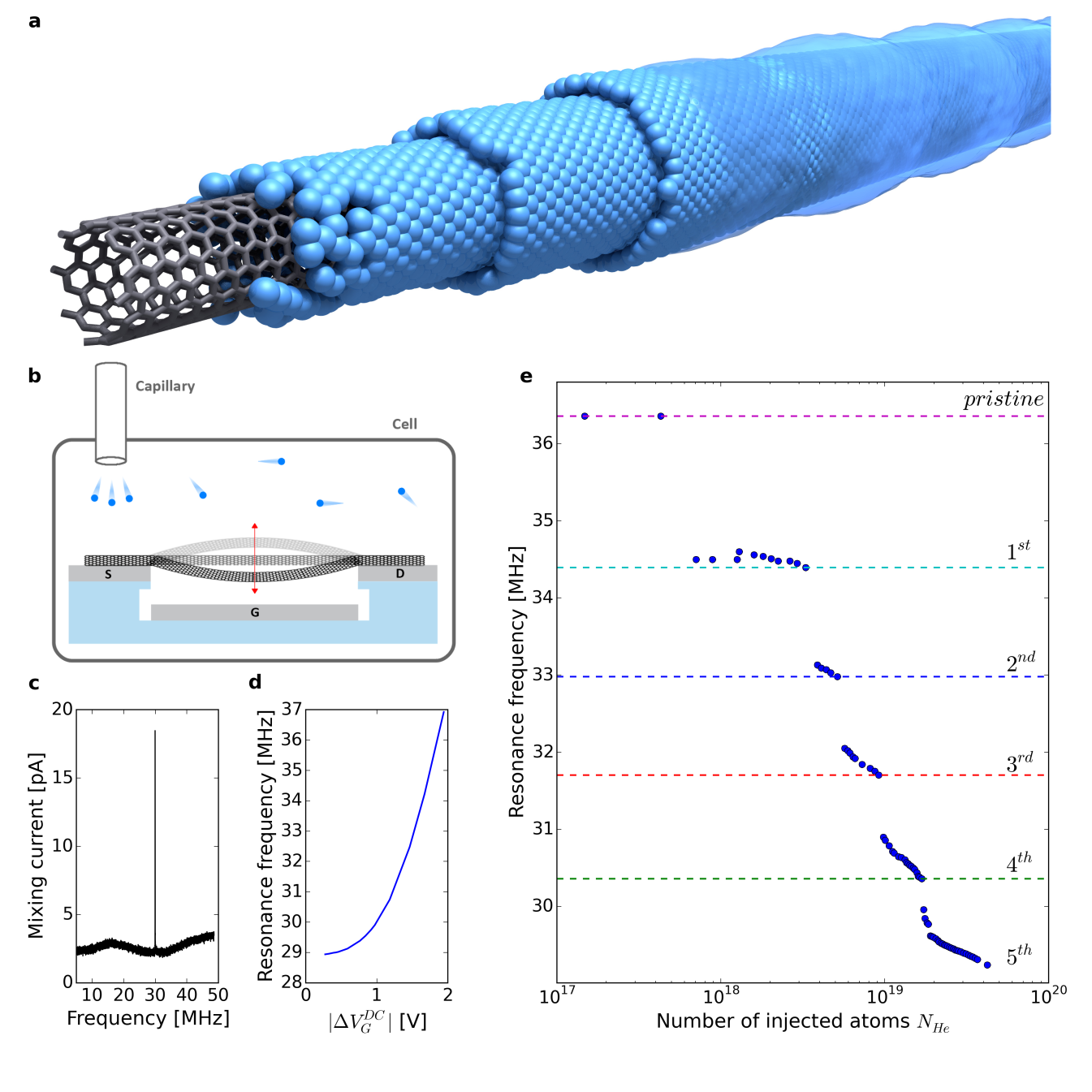}
\caption{{\bf Helium multilayer adsorbed on a nanotube.} {\bf(a)} Graphical representation of the helium multilayer on the nanotube. {\bf(b)} Schematic of the nanotube mechanical resonator immersed in helium vapour. The nanotube is contacted to the platinum electrodes $S$ and $D$ and is capacitively coupled to the electrode $G$. {\bf(c)} Spectrum of the mechanical vibration amplitude as a function of the drive frequency at 0.1~K and $V_\mathrm{G}=1.908$~V. {\bf (d)} Gate voltage dependence of the resonance frequency. The offset due to the work function difference between the nanotube and the gate electrode is $V_\mathrm{G}=0.077$~V. {\bf (e}) Layer-by-layer growth of the helium film. The resonance frequency is measured at 20~mK as a function of the number of helium atoms injected at room temperature into the cell through the capillary. The different steps are associated to the pristine nanotube and the successive adsorption of helium layers; the steps are labeled accordingly and highlighted by the horizontal dashed lines.}
\label{setup}
\end{figure}

Recent advances in nanomechanics have already allowed studies of liquids at small scales~\cite{Favero2015, Kashkanova2017, Carmon2016, Harris2016, Bowen2016} but  the study of the adsorption of superfluid helium on a single carbon nanotube is original (Fig.~\ref{setup}a).
We  fabricate mechanical resonators by suspending a single NT of 3~nm diameter and 1.1~$\mu$m length. This NT is contacted to two platinum electrodes and is capacitively coupled to a gate electrode (Fig.~\ref{setup}b). In order to suppress any surface contamination, we grow the nanotube in the last step of the fabrication process and we anneal it by passing a large current (6~$\mu$A)  through it in the dilution refrigerator at 20~mK prior to the adsorption of helium. This procedure already allowed us to grow xenon monolayer crystals that were commensurate with the nanotube lattice~\cite{Tavernarakis2014}. The nanotube vibrates as a doubly-clamped string. The mechanical vibrations of the fundamental mode of the nanotube are driven capacitively and measured electrically (Fig.~\ref{setup}c)~\cite{McEuen2004, Moser2014,Bonis2018}. The resonance frequency can be tuned by a large amount around 30~MHz when tuning the gate voltage (Figs.~\ref{setup}c and \ref{setup}d). The quality factor deduced from the thermal motion linewidth is $\approx 2 \cdot 10^{5}$. We inject $^4$He from room temperature into the sample cell through a capillary.

A mechanical resonator based on a suspended nanotube is a remarkable sensor of atoms adsorbed on its surface \cite{Chaste2012,Bockrath2008,Cobden2010,Cobden2012}. The mass detection can reach single atom resolution \cite{Chaste2012}. Driving the resonator to detect adsorbed atoms does not affect the dynamics of the adsorbates, since the amplitude of the driven vibrations can be kept smaller than the amplitude of thermal vibrations. The resonance frequency  $f_0=\frac{1}{2\pi}\sqrt{\frac{K}{M}}$ depends on the  ratio of the string elasticity also called spring coefficient $K$  to the effective mass $M$.  The coefficient $K = k_\mathrm{NT}+ k_\mathrm{He}$ is the sum of the respective elasticities of the naked NT- also called pristine -  and of the adsorbed helium.  Similarly the total mass $M = m_\mathrm{NT} + m_\mathrm{He}$ is the sum of the respective masses of the NT and of the adsorbed helium. With helium adsorbed, both $K$ and $M$ increase but their contributions  to the resonance frequency have opposite signs, so that we could distinguish between them by varying the gate voltage. This is a great advantage of our experiment. Previous studies  of two-dimensional helium superfluids  often used torsional oscillators~\cite{Reppy1978,Reppy1996, Saunders2017}. Contrary to our case, these torsional oscillators were macroscopic with not much  possibility to change their resonance frequency so that mass effects were difficult to distinguish from elasticity effects~\cite{Reppy2014}. As we shall see below, we have observed the two effects in two different temperature regimes.

By studying the transport properties of electrons inside the NT, we have demonstrated that our nanotubes are free of disorder. As a function of the gate voltage, the conductance is periodically modulated due to quantum electron interference~\cite{Liang2001}. This periodic modulation would be deteriorated by a tiny amount of disorder (see section I of Supplementary Information).

We  measured the nanotube length $L \simeq 1.1~\mu$m from the characteristic source-drain voltage $V_\mathrm{C}=1.5$~mV of the electron interference pattern. This length is $L=hv_\mathrm{F}/2eV_\mathrm{C}$~\cite{Liang2001} where $v_\mathrm{F}=8\times10^5$~m/s is the Fermi velocity of nanotubes. It is consistent with the width of the trench and we obtained a similar  but less reliable length ($L \approx 1~\mu m$) from the Coulomb blockade measurements of the conductance.

\begin{figure}[h!]
	\centering
	\includegraphics[scale=0.45, trim = 0cm 0cm 0cm 0cm, clip=true]{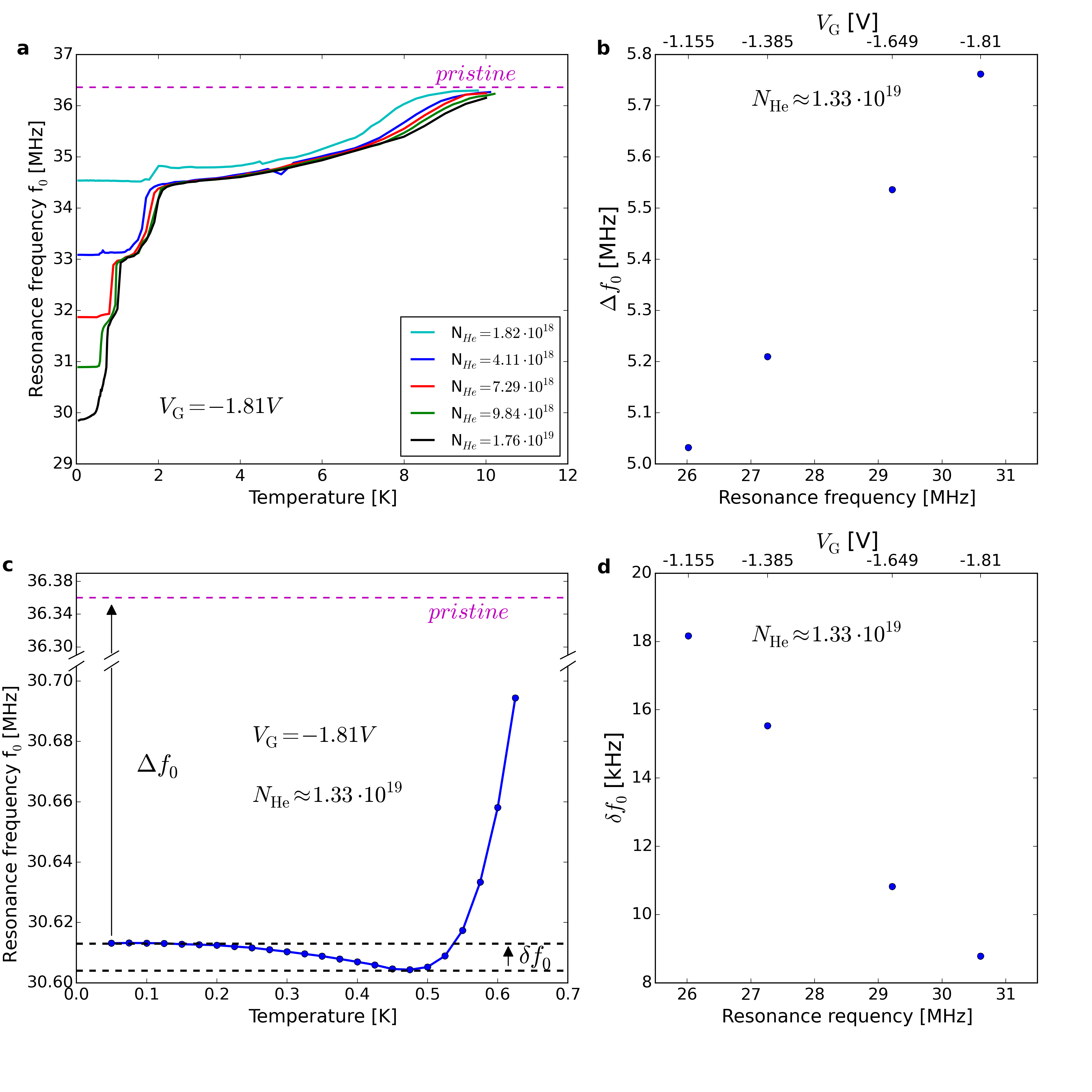}
	\caption{{\bf Layer-by-layer desorption upon increasing temperature.} {\bf(a)} Resonance frequency as a function of temperature. The ''pristine'' is the carbon nanotube without any adsorption. {\bf(b)} Frequency shift $\Delta f_\mathrm{0}$ related to $m_\mathrm{He}$ as a function of gate voltage (upper axis) and resonance frequency (lower axis). As indicated in {\bf c}, $\Delta f_\mathrm{0}$ is the change in resonance frequency between the pristine nanotube and the dressed nanotube at 20 mK. 
{\bf(c)} Temperature dependence of the resonance frequency in the low-temperature regime. {\bf(d)} Frequency shift $\delta f_\mathrm{0}$ related to $k_\mathrm{He}$ as a function of gate voltage (upper axis) and resonance frequency (lower axis). The shift $\delta f_\mathrm{0}$ is represented in {\bf c}. 
}
\label{layering_a}
\end{figure}

\begin{figure}[h!]
	\centering
	\includegraphics[scale=0.45, trim = 0cm 0cm 0cm 0cm, clip=true]{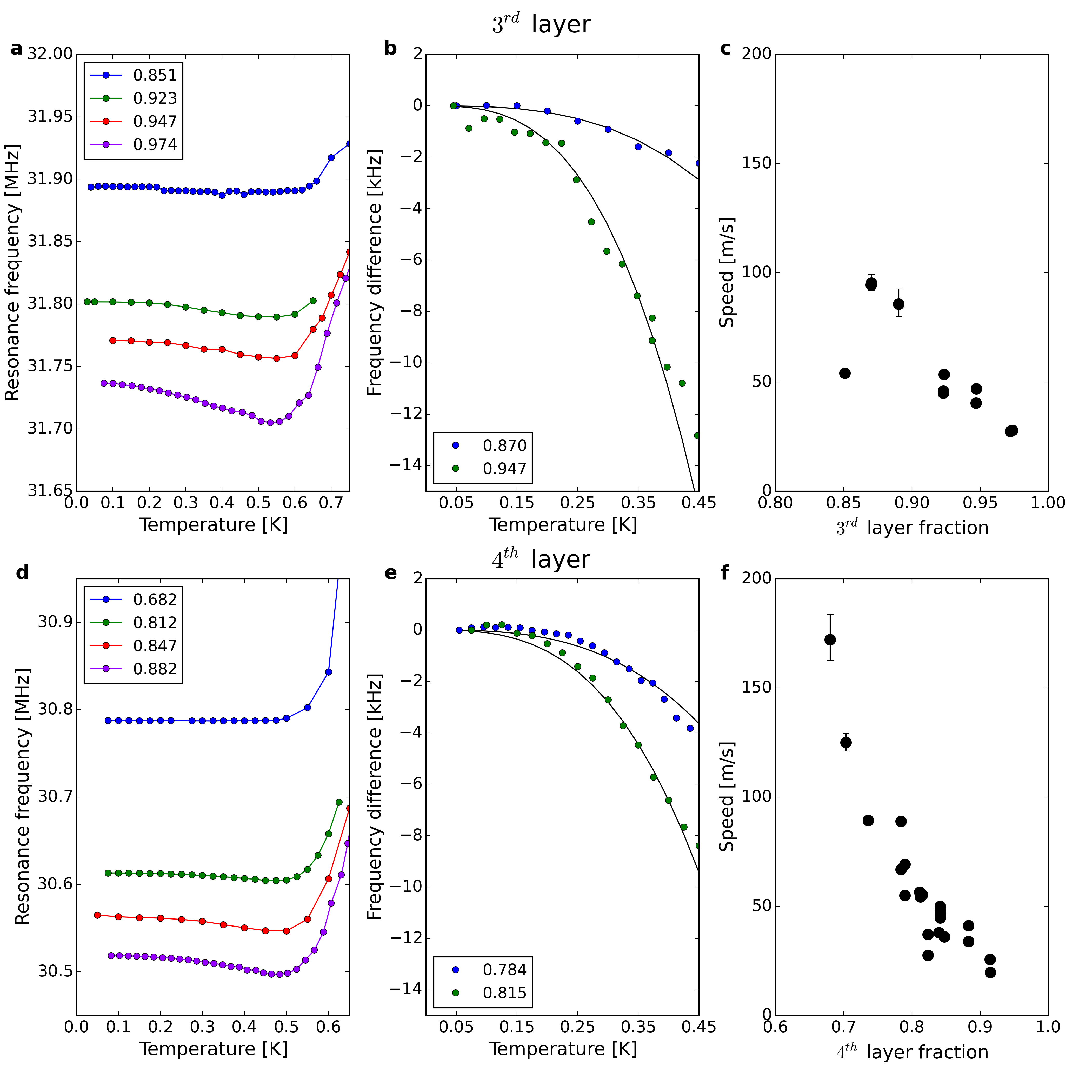}
	\caption{{\bf Mechanical sensing of the third sound.} {\bf(a)} Resonance frequency as a function of temperature in the low-temperature regime for the third layer. The different curves correspond to different layer filling fractions, which are indicated in the legend. {\bf(b)} Measured temperature dependence of the resonance shift (dots). The lines are cubic temperature dependencies, as expected from Eq.~\ref{eq:Df}. {\bf(c)} Third sound speed as a function of the layer filling fraction of the third layer. The error bars are obtained from the comparison between the measured $\delta f_0 (T)$ dependence and Eq.~\ref{eq:Df}. {\bf(d-f)} Same as {\bf a-c} but for the fourth layer.}
\label{layering_b}
\end{figure}

We observe a series of steps when measuring the resonance frequency of the nanotube as a function of the number $N_\mathrm{He}$ of helium atoms injected into the sample chamber (Fig.~\ref{setup}e). We proceed by successive injections of $\approx$ 76~cm$^3$ of helium gas at a pressure set between 0.08 and 3 mbar. When we increase $N_\mathrm{He}$, the vapour pressure $P$ in the surrounding of the nanotube gets larger, but $P$ cannot be quantitatively estimated due to the unknown adsorption of helium atoms onto the cell walls. We attribute the reduction of the resonance frequency $\Delta f_0$ measured in Fig.~\ref{setup}e to the helium mass $m_{He}$ adsorbed on the NT. Each plateau in $f_{0}$ is assigned to a different layer adsorbed on the nanotube. At the completion of each layer, a new layer starts and the resonance frequency jumps down. It means a divergence of the coverage at constant helium pressure, that is, a divergence of the compressibility and a vanishing of the sound speed in the helium layer. This is in agreement with the first-order transition predicted for the layering transition \cite{Clements1996}. The helium density at each discontinuity agrees within 6~\% with the established density of helium adsorbed on graphite (Section~III of Supplementary Information).
During the completion of each layer, helium atoms must be homogeneously distributed over the surface and the chemical potential of the helium vapor needs to increase to equilibrate the increase of the repulsion betwen atoms. This explains the observed finite slope $\partial f_0/\partial N_\mathrm{He}$ of the plateaus (Fig.~\ref{setup}e).

Figure~\ref{layering_a}a shows the layer-by-layer desorption of helium from the surface of the nanotube, but now as a function of temperature up to 10~K and for 5 successive amounts of helium in the cell. The frequency of the steps are similar to those measured while varying $N_\mathrm{He}$ at constant temperature (Fig.~\ref{setup}e). One temperature sweep takes a few days to ensure the reproducibility of measurements. The layering transitions are rounded, probably because the layering transitions have a critical point of order 1~K (see Ref. \cite{Ramesh1984}). By varying the gate voltage, consequently the spring constant, we have confirmed that the variation of the resonance frequency $f_0$ in Fig.~\ref{layering_a}a is due to the change in the resonator mass. Indeed the large frequency change $\Delta f_0$  is an increasing function of $f_0$ (Fig.~\ref{layering_a}b and  Supplementary Section IV).

The first and second layers show some peculiar structures which would need further study to be compared with predictions by Boronat et al. (see Ref.~\cite{Boronat2012} and Section~VI of Supplementary Information).  We thus focus on the third and fourth layers. By looking more precisely at the frequency variation below 0.7~K, we discovered the existence of a small minimum near 0.5~K (see Fig.~\ref{layering_a}c). The rise of $f_0$ above $\sim$0.5~K in Fig.~\ref{layering_a}c  signals the desorption of helium atoms from the nanotube. Figure~\ref{layering_a}d shows the dependence of the depth $\delta f_0$ of this small minimum on the gate voltage. Contrary to the case of the large frequency shift $\Delta f_0$ of Fig.~\ref{layering_a}b, the slope is negative (see Fig.~\ref{layering_a}d and Supplementary Section IV). It means that this small minimum is due the helium contribution  to the total spring constant of the resonator.

We attribute it to the temperature variation of the helium surface tension, that is, to the growing entropy of the helium film~\cite{Atkins1965}. It is due to the third sound states in the case of thin superfluid films~\cite{Atkins1959}. These waves are predicted to be longitudinal with no sizeable displacement of helium atoms in the direction perpendicular to the surface, except for densities near the layering transition~\cite{Clements1994}. The third sound has a linear energy dispersion given by the speed $c$. The resulting dependence of $\delta f_0$ on temperature $T$ is
\begin{equation}
\delta f_0 (T)=-0.074\frac{1}{m_\mathrm{NT}f_0}\frac{r_\mathrm{He}}{L}\frac{(k_\mathrm{B}T)^3}{(\hbar c)^2}
\label{eq:Df}
\end{equation}
where $m_\mathrm{NT}$ is the mass of the pristine nanotube, $L$ the suspended nanotube length, $k_\mathrm{B}$ the Boltzmann constant, and $\hbar$ the reduced Planck constant. See  Section~V of the Supplementary Information for a detailed derivation.

Figures~\ref{layering_b}b and \ref{layering_b}e show a very good agreement with the $T^3$ law  of Eq.~(\ref{eq:Df}) from which we could extract the sound velocity $c$ (Figs.~\ref{layering_b}c and f). The layer fraction are computed by comparing the resonance frequency at 20 mK to the frequency of the jump indicated by the dashed line on Fig.~\ref{setup}e. We notice a large jump in sound velocity from the end of the 3$^{rd}$ layer (30~m/s) to the beginning of the 4$^{th}$ layer (210 m/s), in agreement with the predictions of a first order transition by Clements et al. \cite{Clements1993,Clements1996}. Although we found no evidence of ripplon excitations in our system, studying thicker helium layers may lead to their observation in future work.

Our work shows that helium layers  adsorbed on carbon nanotubes are of unprecedented quality. This system is of great interest for the study of different quantum phenomena. The cylindrical boundary condition of the superfluid imposes quantized translational velocities of a vortex around the circumference~\cite{Guenther2017}. This boundary condition also modifies the vortex-vortex interaction energy~\cite{Guenther2017}, so that it might alter the Berezinskii-Kosterlitz-Thouless topological phase transition. According to Gordillo et al.~\cite{Gordillo2011}, helium monolayers adsorbed on nanotubes may become supersolid at low temperature. In summary, carbon nanotubes could be used for original studies of helium-based quantum phenomena.


\textbf{Acknowledgments} We thank N. Guenther, P. Massignan, A. Fetter, and G. A. Williams for fruitful discussions. This work is supported by the ERC advanced grant 692876, the Foundation Cellex, the CERCA Programme, AGAUR, Severo Ochoa (SEV-2015-0522), the grant FIS2015-69831-P of MINECO, and the Fondo Europeo de Desarrollo Regional (FEDER). M.C.G. acknowledges partial financial support from the MINECO (Spanish Ministry of Economy) Grant No. FIS2017-84114-C2-2-P. M.C.G. also acknowledges the use of the C3UPO computer facilities at the Universidad Pablo de Olavide. J.B. acknowledges financial support from MINECO Grants FIS2014-56257-C2-1-P and FIS2017-84114-C2-1-P.

\section*{References}


\end{document}


\title{Supplementary Information\\Layering transition in superfluid helium adsorbed on a carbon nanotube mechanical resonator}

\author{Adrien Noury}
\author{Jorge Vergara-Cruz}
\author{Pascal Morfin}
\author{Bernard Pla\c{c}ais}
\author{Maria C. Gordillo}
\author{Jordi Boronat}
\author{S\'{e}bastien Balibar}
\author{Adrian Bachtold}

\maketitle


\section{Quantum electron transport measurements}\label{electrontransport}
\begin{figure}[t]
\includegraphics[width=12cm]{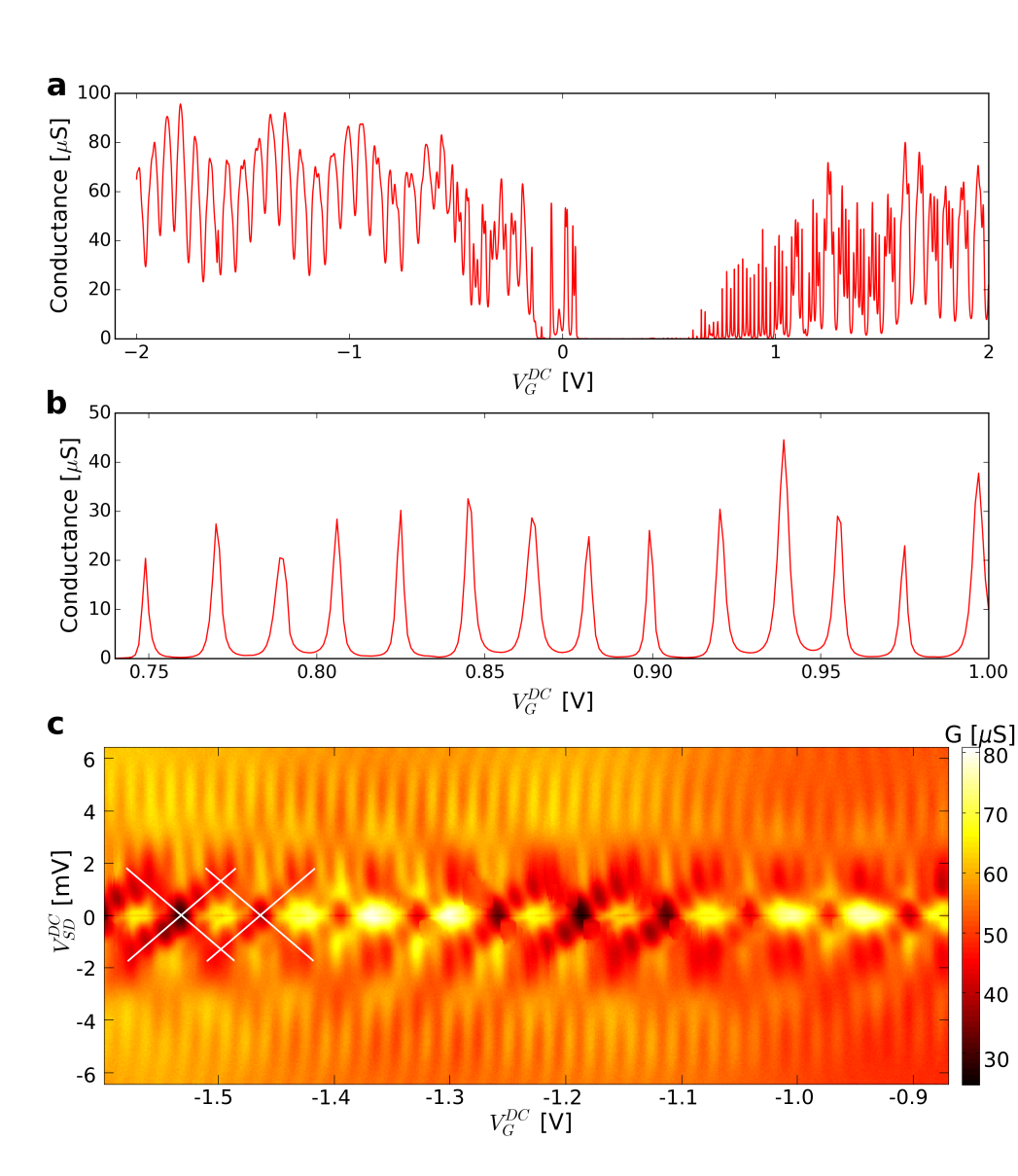}
\caption{{\bf Electron transport measurements of the nanotube discussed in the main text.} {\bf (a,b)} Conductance of the nanotube as a function of gate voltage measured at the base temperature of the cryostat over two different gate voltage ranges. {\bf (c)} Differential conductance as a function of $V_\mathrm{SD}^\mathrm{DC}$ and $V_\mathrm{G}^\mathrm{DC}$ at the base temperature of the cryostat. The intersection of the white lines at $V_\mathrm{C}=1.5$~mV corresponds to the characteristic voltage bias of the electron interference }
\label{conductance}
\end{figure}

Figure~\ref{conductance}a shows that the electrical characteristics of the nanotube studied in this work is typical of ultraclean nanotubes \cite{Moser2014}. Near $V_\mathrm{G}^\mathrm{DC}=0$~V, the conductance is suppressed to zero due to the small energy gap of the nanotube~\cite{Laird2015}. For positive $V_\mathrm{G}^\mathrm{DC}$ values, $p-n$ junctions are formed near the metal electrodes. It creates Coulomb blockade peaks in the conductance (Fig.~\ref{conductance}b). For negative $V_\mathrm{G}^\mathrm{DC}$, the nanotube is $p$-doped along the whole tube, resulting in a larger conductance approaching the quantum conductance $4e^2/h$ of small-gap nanotubes. In this regime, the conductance is modulated due to quantum electron interference~\cite{Liang2001}. Here, $e$ is the charge of the electron and $h$ is the Planck constant.

Figure~\ref{conductance}c demonstrates that the nanotube is of high quality, since the modulation of the conductance is periodic over a large range of $V_\mathrm{G}^\mathrm{DC}$, and since this periodic modulation due to electron interference would be deteriorated by a tiny amount of disorder.

We quantify the nanotube length $L=1.1~\mu$m from the characteristic voltage bias $V_\mathrm{C}=1.5$~mV of the electron interference pattern shown in Fig.~\ref{conductance}c using $L=hv_\mathrm{F}/2eV_\mathrm{C}$~\cite{Liang2001}. Here, $v_\mathrm{F}=8\times10^5$~m/s is the Fermi velocity of nanotubes. This length is consistent with the width of the trench. We also obtain a similar length from the Coulomb blockade measurements in Fig.~\ref{conductance}b. From the separation $\Delta V_\mathrm{G}^\mathrm{DC}=18$~mV between two
conductance peaks, we obtain the nanotube-gate capacitance $C_\mathrm{G}=e/\Delta V_\mathrm{G}^\mathrm{DC}=8.9\times10^{-18}$~F. We get the length $L\simeq 1~\mu$m of the suspended nanotube using
\begin{equation}
C_\mathrm{G}=\frac{2\pi\epsilon_{0}L}{\ln\left(\frac{2d}{r}\right)},
\label{CG}
\end{equation}
where $\epsilon_{0}$ is the vacuum permittivity, $d=350$~nm the separation between the nanotube and the gate electrode, and $r$ the nanotube radius. This estimation of the length is less reliable than the previous one because Eq.~\ref{CG} does not take into account the screening of the electric field between the nanotube and the gate electrode by the source and the drain electrodes.

\section{Transduction of displacement into current}\label{vibration}
We measure the mechanical vibrations of the nanotube with the two-source technique~\cite{Sazanova2004,Bonis2018}. Displacement modulations result in current modulations by applying an input oscillating voltage with amplitude $V_\mathrm{SD}^\mathrm{AC}$ across the nanotube. We assume that the resonance used in the measurements of the main text corresponds to the fundamental eigenmode polarized in the direction perpendicular to the surface of the gate electrode, which is a good assumption since the signal of the driven vibrations of this resonance is much larger than the signal of the other resonances. The current $\delta I$ at the frequency close to the difference between the
mode eigenfrequency and the frequency of the source-drain voltage
is
\begin{eqnarray}
&& \delta I=\beta\delta z ,
\label{eq:Iz}\\
&& \beta=\frac{1}{2}\frac{dG}{dV_\mathrm{G}}V_\mathrm{G}^\mathrm{DC}V_\mathrm{SD}^\mathrm{AC}\frac{C_\mathrm{G}^{\prime}}{C_\mathrm{G}}.
\label{eq:beta}
\end{eqnarray}
Here, $\delta z$ is the displacement of the nanotube, $dG/dV_\mathrm{G}$ is the derivative of the conductance with respect to the gate voltage, $V_\mathrm{G}^\mathrm{DC}$ is the
static gate voltage, and $C_\mathrm{G}^{\prime}$ is the derivative of $C_\mathrm{G}$ with respect to $z$. We quantify $C_\mathrm{G}^{\prime}=4.1\times10^{-12}~\textrm{F/m}$ using the relation
$C_\mathrm{G}^{\prime}=\frac{C_\mathrm{G}}{d\ln(2d/r)}$.

\begin{figure}[t]
\includegraphics[width=12cm]{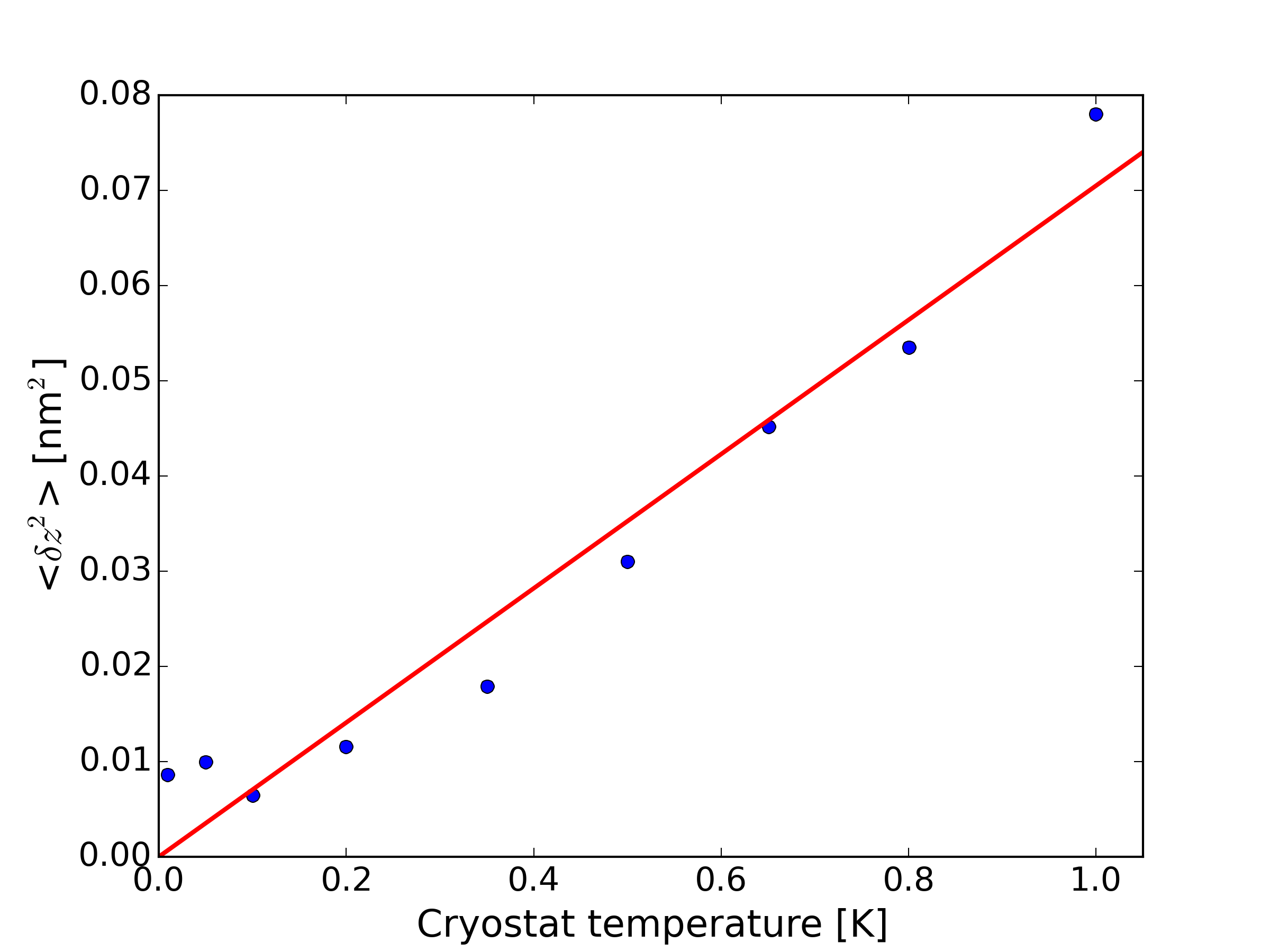}
\caption{{\bf Variance of the mechanical displacement as a function of temperature.} Measurements are carried out on the nanotube discussed in the main text.}
\label{calibration}
\end{figure}

We estimate the effective mass $m=3.5$~ag from the measurement of the variance of the displacement noise $\langle\delta z^{2}\rangle$ as a function of temperature $T$ (Fig.~\ref{calibration}). The displacement noise of the nanotube is measured with the electrical method described in Ref.~\cite{Bonis2018} and using Eqs.~\ref{eq:Iz} and ~\ref{eq:beta}. We compare the measured slope of $\langle\delta z^{2}\rangle$ as a function of $T$ to the slope expected from the equipartition theorem, which reads $m\omega_{0}^2\langle\delta z^{2}\rangle=k_{b}T$. Here, $\omega_{0}/2\pi$ is the resonance frequency of the eigenmode and $k_{b}$ is the Boltzmann constant. This mass is consistent with the mass of a $\sim1.1~\mu$m long nanotube.

We evaluate the nanotube radius $r\simeq 1.5$~nm from the effective mass using the relation
\begin{equation}
m=\frac{1}{2}\left(2m_\mathrm{C}\times\frac{2\pi r\times L}{A_\mathrm{h}}\right),
\label{meff}
\end{equation}
where $m_\mathrm{C}=2\times10^{-26}$~kg is the mass of a carbon atom and $A_\mathrm{h}=5.2\times10^{-20}$~m$^{2}$ is the surface area of a hexagon in the honeycomb lattice of graphene. The coefficient $\frac{1}{2}$ on the right-hand side of Eq.~\ref{meff} comes from the normalisation of the mass of the resonator due to the shape of the eigenmode. We assume here that the modal shape is $\phi(x)=\mathrm{cos}(\pi x/L)$, a good approximation for the shape of a beam under tension.

\begin{table*}[b]
\begin{center}
 \begin{tabular}{|p{4cm}|p{3cm}|p{3cm}|p{3cm}|}
 \hline
\multicolumn{2}{|c|}{ } & \multicolumn{2}{|c|}{{\bf Areal density (atoms/nm$^2$)}} \\
\hline
   & $f_0$ (MHz) & nanotube & graphite \\
 \hline\hline
 Pristine substrate  & 36.34  &0  &0  \\
 \hline
 First layer completed & 34.41  &11.0  &11.4 \\
\hline
Second layer completed  & 32.97  & 8.1 &8.6 \\
 \hline
Third layer completed  & 31.69  & 7.2 &7.6\\
 \hline
Fourth layer completed  & 30.39  & 7.3 &7.6\\

\hline

\end{tabular}
\caption{Areal density of completed helium layers on the nanotube. The density of helium adsorbed on graphite is also shown~\cite{Crowell1996}.}
\label{table:tb}
\end{center}
\end{table*}

\section{Density of helium layers}\label{density}
We can reliably quantify the ratio between the number $N_{\rm He}$ of adsorbed helium atoms and the number $N_{\rm C}$ of carbon atoms at the surface of the nanotube from the measurement of the resonance frequency~\cite{Wang2010,Tavernarakis2014}. This ratio, called coverage, reads
\begin{equation}
 \phi=\frac{N_{\rm He}}{N_{\rm C}}=\frac{m_{\rm C}}{m_{\rm He}}\Bigg[\Bigg(\frac{f_0^{\rm NT}}{f_0^{\rm NT+He}}\Bigg)^2-1\Bigg],
\label{eq:coverage}
 \end{equation}
where $m_{\rm C}$ and $m_{\rm He}$ are the atomic masses of carbon and helium atoms, respectively. Here, $f_0^{\rm NT+He}$ is the
resonance frequency of the nanotube covered by helium atoms, and $f_0^{\rm NT}$ is the resonance frequency of the pristine nanotube without any adsorbed helium atoms. The coverage of noble gas atoms adsorbed on nanotube resonators can be successfully quantified with Eq.~\ref{eq:coverage} because these adsorbed atoms increase the mass of the resonator but do not modify its spring constant~\cite{Wang2010,Tavernarakis2014}. Indeed, the interaction between noble gas atoms is much weaker than that between the carbon atoms of the nanotube.

Table~\ref{table:tb} shows the areal density of the different completed helium layers. The areal density is estimated from the coverage and taking into account the cylindrical geometry that normalizes the density by the factor $\frac{r}{r+i\cdot d_{\rm l}}$. Here, $r$ is the nanotube radius, $i$ is the layer number, and $d_{\rm l}$ is the layer-layer separation. The characterisation of the nanotube in the previous section indicates that $r=1.5$~nm. The separation between layers is taken equal to $d_{\rm l}=0.27$~nm, the separation between the first helium layer and graphite~\cite{Corboz2008}. Table~\ref{table:tb} shows that the estimated areal densities are remarkably close to the values measured with helium adsorbed on graphite~\cite{Crowell1996}.

\section{Resonance frequency shift -- A mass effect or a spring effect?}\label{massspring}
We look at the $f_0$ dependence of two different frequency shifts, namely $\Delta f_0$  and $\delta f_0$ (see Fig.~2 in the article), which show opposite behavior. The resonance frequency of the nanotube covered by helium depends on three parameters: the helium pressure in the cell $P_\mathrm{He}$, the temperature $T$, and the gate voltage $V_\mathrm{G}^\mathrm{DC}$. The resonance frequency of the bare nanotube depends only on the DC gate voltage $V_\mathrm{G}^\mathrm{DC}$ as
\begin{equation}
f_0 (V_\mathrm{G}^\mathrm{DC}) = \frac {1}{2\pi}\sqrt{\frac{k_\mathrm{NT}(V_\mathrm{G}^\mathrm{DC})}{m_\mathrm{NT}}},
\label{eq:f0}
\end{equation}
where the nanotube stiffness $k_\mathrm{NT}$ can be changed by tuning the gate-voltage $V_\mathrm{G}^\mathrm{DC}$, in contrast to the nanotube mass $m_\mathrm{NT}$. We measure that $f_0 (V_\mathrm{G}^\mathrm{DC})$ is temperature independent at a fixed $V_\mathrm{G}^\mathrm{DC}$.  Straining the nanotube with gate voltage is very efficient, since $f_0(V_\mathrm{G}^\mathrm{DC})$ can be varied by $\sim 16\%$ in Figs. 2b and 2d of the article. Due to the large nanotube stiffness, the longitudinal deformation of the nanotube is minute and the helium film structure also. Furthermore, the surface density of helium on the nanotube is fixed by the temperature and the helium pressure in the cell.

The resonance frequency of the nanotube coated by a thin helium film is:
\begin{equation}
f_1 = \frac {1}{2\pi}\sqrt{\frac{k_{NT}+k_{He}}{m_\mathrm{NT}+m_\mathrm{He}}} \approx f_0 (1-\frac{m_\mathrm{He}}{2m_\mathrm{NT}}+\frac{k_{He}}{2k_{NT}} )
\label{eq:f1}
\end{equation}
High-frequency nanotube mechanical resonators are highly sensitive to helium coating. Loading the nanotube with a single helium layer gives rise to a mass increment $m_\mathrm{He} / m_\mathrm{NT} \lesssim 10\% $  and a decrease of the resonance frequency  $(f_1-f_0 ) / f_0  \approx - m_\mathrm{He} /2m_\mathrm{NT} \lesssim -5\%$. A second effect of the helium coverage is the modification of the nanotube spring constant due to a modification of the surface tension. The effect is extremely small because of the weak He-He interaction compared to the covalent C-C interaction, but still measurable thanks to the remarkable sensitivity of high-Q nanotube resonators. By studying the $V_\mathrm{G}^\mathrm{DC}$ dependence of the frequency shifts, it is possible to distinguish the two contributions. Indeed, as a function of the helium pressure and the temperature, the frequency shift is expressed as
\begin{eqnarray}
&& f_0 - f_1  \approx f_0(V_\mathrm{G}^\mathrm{DC})\left( \frac{m_\mathrm{He}(P_\mathrm{He},T)}{2m_\mathrm{NT}} - \frac{k_\mathrm{He}(P_\mathrm{He},T)}{2k_{NT}(V_\mathrm{G}^\mathrm{DC})}\right),
\label{eq:f1-f0}\\
&& f_0 - f_1  \approx \left( \frac{1}{2m_\mathrm{NT}}\right )  \left( f_0(V_\mathrm{G}^\mathrm{DC}) \times m_\mathrm{He}(P_\mathrm{He},T) - \frac{k_\mathrm{He}(P_\mathrm{He},T)}{4\pi^2 f_{0}(V_\mathrm{G}^\mathrm{DC})}\right).
\label{eq:f1-f0bis}
\end{eqnarray}
Mass effects increase proportionally to $f_0$ whereas stiffness effects decrease proportionally to $(1/f_0)$. The former depends on the adsorbed helium mass, consequently on the gas pressure, on temperature, possibly also on changes in superfluid mass fraction. The later addresses the physics of the helium surface tension which carries additional signatures of the layering transition.  Note that a $V_\mathrm{G}^\mathrm{DC}$-dependence of $k_\mathrm{He}$ and $m_\mathrm{He}$ would intervene as a second order correction in the developments, which justifies the above variable separation.

Let us now look at the $f_0$ dependence of the frequency shift $\Delta f_0$ that is indicated on Fig.~2c of the article. We expect it to be a mass effect due to the helium evaporation between 10 mK and 10K so that $\Delta f_0$ should obey the equation:
\begin{equation}
\Delta f_0  \approx  \frac{f_0  (V_\mathrm{G}^\mathrm{DC} ) m_\mathrm{He}(P_\mathrm{He},10~\mathrm{mK}) }{2m_\mathrm{NT}}.
\label{eq:masseffect}
\end{equation}
$\Delta f_0$ should be an increasing function of $f_0$, in agreement with the measurements on Fig.~2b of the article.

On the opposite, Fig.~2d of the article shows that $\delta f_0$ is a decreasing function of $f_0$ which demonstrates that this slight frequency dip is due to some change in the elastic constant $k_\mathrm{He}$. Indeed it writes now:
\begin{equation}
\delta f_0  \approx  - \frac{\delta k_\mathrm{He}(P_\mathrm{He})}{8\pi^2  m_\mathrm{NT} f_{0}(V_\mathrm{G}^\mathrm{DC})}
\label{eq:tensioneffect}
\end{equation}
where $\delta k_\mathrm{He}$ is the change in the helium surface tension from 10 mK to the temperature of the minimum (about 0.5K, see Fig.~2c of the article). Here, $\delta k_\mathrm{He}$ is negative and we attribute it to the increasing entropy of the helium film. In this case, one expects that $\delta f_0$ decreases with $f_0$, which is confirmed by our measurements (see Fig.~2d of the article). We cannot attribute it to some superfluid-normal transition in the film because this would be a mass effect with the opposite dependence on $f_0$. As shown in the next section, we have calculated the temperature dependence of the helium surface tension, which can describe our measurements.

\section{Surface tension -- Thermally excited third sound states}\label{surfacetension}

We show in the main text that the measured spring constant of the nanotube covered by superfluid helium is temperature dependent. In this section, we relate this observation to the change of the surface tension of the superfluid due to thermally excited third sound states.

The surface tension $\gamma$ is the free energy of the superfluid surface per unit area. When varying the superfluid surface area by $\delta A$, the energy changes as
\begin{equation}
\delta E=\gamma \cdot \delta A.
\label{STdefinition}
\end{equation}
The surface tension contributes to the spring constant of the resonator. Any small deviation from the equilibrium position of the superfluid leads to a spring force. The modulation $\delta A$ is related to the modulation of the resonator length $\delta l$ when the nanotube is vibrating. The modulation is $\delta A=2\pi r_\mathrm{He} \cdot\delta l$ where $r_\mathrm{He}$ is the radius of the surface of the helium superfluid covering the nanotube. In order to relate $\delta l$ to the displacement $\delta z$ of the nanotube resonator, we consider the fundamental mode of a doubly-clamped nanotube string. The transverse displacement of the resonator along its axis $x$ is given by $Z(x,t)=\delta z(t) \cdot\phi (x)= \delta z(t) \cdot \mathrm{cos}(\pi x/L)$ with $\phi (\pm L/2)=0$ the boundary conditions at the clamping points. When the nanotube moves by $\delta z$, the total length becomes
\begin{equation}
L+\delta l=\int_{-L/2}^{+L/2} dx\sqrt{1 + \left( \frac{\partial Z}{\partial x} \right) ^2}\simeq L+\frac{1}{2}\int_{-L/2}^{+L/2} dx\left(\frac{\partial Z}{\partial x}\right)^2=L+\frac{\pi^2}{4L}\delta z^2.
\label{dxdl0}
\end{equation}
As a result, the energy associated to the surface tension of the superfluid film is the energy of a harmonic oscillator $\delta E=\frac{1}{2}k_\mathrm{He}\delta z^2$ with spring constant
\begin{equation}
k_\mathrm{He}=\gamma \pi^3 \frac{r_\mathrm{He}}{L}.
\label{kdEdA}
\end{equation}
The restoring force of the helium film acts in parallel to the restoring force of the pristine carbon nanotube $k_\mathrm{NT}$, so that the total spring constant is $k_\mathrm{total}=k_\mathrm{He}+k_\mathrm{NT}$. It is important to emphasize that the helium film contributes weakly to the total spring constant, since the interaction between noble gas atoms is 2 orders of magnitude weaker than that of covalent C-C bonds, as demonstrated experimentally in Refs.~\cite{Wang2010,Tavernarakis2014}. For this reason, any change of the surface tension leads to a minuscule change of the resonance frequency of the resonator, as observed in our experiments discussed in the main text.

It is interesting to compare the typical elongation $\delta l$ in our experiments with the separation between two helium atoms. The modulation of the elongation is small so that the number of helium atoms adsorbed on the suspended nanotube remains constant. Indeed, the amplitude of the thermal vibrations is $\leq 300$~pm below $T=1$~K so that the associated nanotube elongation is $\delta l \leq 0.2$~pm using Eq.~\ref{dxdl0}. For comparison, the separation between helium atoms in thin films is typically 0.3~nm.

The temperature dependence of the surface tension arises from the change of the free energy of the superfluid, that is, from the thermal excitation of two-dimensional third sound states. To compute $\gamma(T)$, we follow the calculations of Atkins using third sound states instead of the surface tension waves in three-dimensional superfluid helium~\cite{Atkins1953}. Below 1~K, the third sound dispersion is given by
\begin{equation}
\omega=ck
\label{disp}
\end{equation}
with the angular frequency $\omega$, the superfluid velocity $c$, and the wavevector strength $k$ ~\cite{Rutledge1978}. In order to calculate the density of states, we first count the number $N_\mathrm{st}$ of states within the surface $\pi k^2$
\begin{equation}
N_\mathrm{st}=\frac{L^2_0}{(2\pi)^2}\pi k^2
\label{Ns}
\end{equation}
with $L^2_0$ the surface area. The density of states is given by the number of states within the frequency $\delta \omega$ and per unit of surface,
\begin{equation}
\frac{g(\omega)\delta \omega}{L^2_0}=\frac{1}{L^2_0}\delta N_\mathrm{st}
=\frac{1}{L^2_0}\frac{dN_\mathrm{st}}{dk}(\frac{d\omega}{dk})^{-1}\delta \omega=\frac{\omega}{2\pi c^2}\delta \omega
\label{DOS2d}
\end{equation}
The internal energy of thermally excited third wave states per unit surface is then
\begin{equation}
U_\mathrm{th}=\int_{0}^{\infty}\frac{\hbar \omega}{\mathrm{exp}(\hbar \omega/k_\mathrm{B}T)-1}g(\omega)d\omega
\label{Uth}
=\frac{1}{2\pi c^2}\frac{(k_\mathrm{B}T)^3}{\hbar^2}\int_{0}^{\infty}\frac{x^2}{\mathrm{exp}(x)-1}dx
=0.38\frac{(k_\mathrm{B}T)^3}{(\hbar c)^2}.
\end{equation}
When the internal energy varies as $U\propto T^\lambda$, the free energy scales as $F=-\frac{1}{\lambda-1}U$ from simple thermodynamics~\cite{Atkins1953}. As a result, the variation of the free energy per unit surface when increasing the temperature from 0~K is
\begin{equation}
\Delta\gamma(T)=U_\mathrm{th}-TS=-\frac{1}{2}U_\mathrm{th}.
\end{equation}
Here, $S$ is the entropy per unit surface. As a result, we obtain
\begin{equation}
\Delta\gamma(T)=-0.19\frac{(k_\mathrm{B}T)^3}{(\hbar c)^2}.
\label{DG2d}
\end{equation}
The change in free energy is related to the change in spring constant using Eq.~\ref{kdEdA}
\begin{equation}
\delta k_\mathrm{He}(T)=-5.89 \frac{r_\mathrm{He}}{L}\frac{(k_\mathrm{B}T)^3}{(\hbar c)^2},
\label{DkT}
\end{equation}
When increasing the temperature from 0~K, the resonance frequency $f_0$ of the nanotube string covered by the helium film is expected to decrease as
\begin{equation}
\delta f_0 (T)=\frac{1}{2}\frac{\delta k_\mathrm{He}(T)}{k_\mathrm{NT}}f_0
=-0.074\frac{1}{m_\mathrm{NT}f_0}\frac{r_\mathrm{He}}{L}\frac{(k_\mathrm{B}T)^3}{(\hbar c)^2}.
\label{Df2d}
\end{equation}

\section{Theoretical calculations}\label{linewidth}
\begin{figure}[h!]
	\label{first_layer}
	\centering
	\includegraphics[width=8cm]{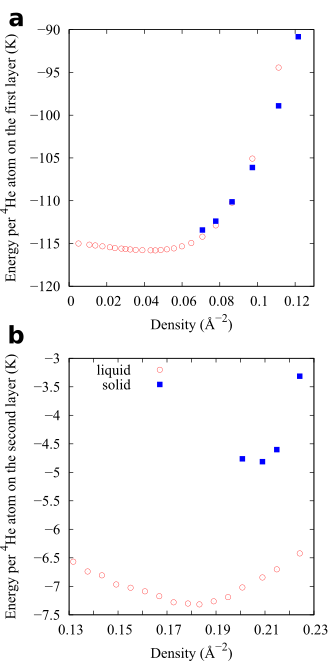}
	\caption{{\bf Calculated energy per atom in the first layer and the second layer} {\bf (a)} First layer. {\bf (b)} Second layer. Energy per atom for both a liquid (red circles) and a solid (blue squares).}
\label{theory}
\end{figure}

Figures~\ref{theory}a,b show the energy per $^{4}$He atom as a function of density for the first and second layer, respectively. Calculations are performed using diffusion Monte Carlo method, as described in ref~\cite{Gordillo2012}. The calculations are performed for a nanotube radius of 1.43 nm.

On Fig.~\ref{theory}a is displayed the energy of atoms in both a liquid (red) and solid (blue) configuration when the nanotube is covered by one layer. The ground state of the system corresponds to a liquid of density $0.0432 \pm 0.003~\mathrm{\AA}^{-2}$. From a density of $0.087 \pm 0.005~\mathrm{\AA}^{-2}$ up, the solid structure is more stable than the liquid. We then conclude that at the second layer promotion (density of $ \approx 0.12~\mathrm{\AA}^{-2}$) the first layer is solid. This is consistent with measurements and calculations on helium adsorbed on graphite~\cite{Crowell1996, Gordillo1998}.

On Fig.~\ref{theory}b is displayed the energy of the atoms in the second layer. The ground state of the system always corresponds to a liquid. The result of these calculations is different from that of previous works on helium adsorbed on graphite, where the second layer at completion is solid~\cite{Crowell1996,Gordillo1998}. More work has to be carried out in order to be able to make a firm conclusion on the phase of the second layer adsorbed on nanotubes. For this, a more advanced model can be used in order to take into account for instance the corrugation of the nanotube.